\documentclass[aps,prl,superscriptaddress,twocolumn,longbibliography]{revtex4-1}
\usepackage{bbm}
\usepackage{graphicx}% Include figure files
\usepackage{dcolumn}% Align table columns on decimal point
\usepackage{bm}% bold math
\usepackage{subfigure}
\usepackage{amsmath}
\usepackage{feynmf}
\usepackage{hyperref}

\usepackage{attachfile}

\newcommand{\bk}{\boldsymbol k}

\newcommand{\bq}{\boldsymbol q}

\newcommand{\bd}{\boldsymbol d}
\newcommand{\bE}{\boldsymbol E}
\newcommand{\bG}{\boldsymbol G}
\newcommand{\bK}{\boldsymbol K}
\newcommand{\bQ}{\boldsymbol Q}

\usepackage{times}

\begin{document}

\title{Nonlinear Hall effect in two-dimensional class AI metals}

\author{Zi-Shan Liao}
\affiliation{School of Physics, Sun Yat-Sen University, Guangzhou 510275, China}
\author{Hong-Hao Zhang}
\affiliation{School of Physics, Sun Yat-Sen University, Guangzhou 510275, China}
\author{Zhongbo Yan}
\email{yanzhb5@mail.sysu.edu.cn}
\affiliation{School of Physics, Sun Yat-Sen University, Guangzhou 510275, China}

\date{\today}

\begin{abstract}
In a time-reversal invariant system, while the anomalous Hall effect
identically vanishes in the linear response regime due to the constraint of time-reversal
symmetry on the distribution of Berry curvature, a nonlinear Hall effect can emerge in the second-order response regime
if the inversion symmetry is broken to allow a nonzero Berry curvature dipole (BCD) on the Fermi surface.
In this work, we study the nonlinear Hall effect of the BCD origin in two-dimensional doped insulators and semimetals
belonging to the symmetry class AI which has spinless time-reversal symmetry.
Despite that the class AI does not host any strong topological insulator phase
in two dimensions, we find that they can still be classified as topologically obstructed insulators and
trivial insulators if putting certain constraint on the Hamiltonians. When the insulator gets closer to the phase boundary of the
two distinct phases, we find that the BCDs  will become more prominent if the doping level is located near the band edge. Moreover,
when the insulator undergoes a phase transition between the two distinct phases, we find that
the  BCDs will dramatically change their signs. For the semimetals without inversion symmetry,
we find that the BCDs will sharply reverse their signs when the doping level crosses
the Dirac points. With the shift of the locations of Dirac points in energy, the critical doping level
at which the BCDs sharply reverse their signs will accordingly change. Our study reveals
that class AI materials can also have interesting geometrical and topological properties, and
remarkable nonlinear Hall effect can also appear in this class of materials even though the
spin-orbit coupling is negligible. Our findings broaden the scope of materials to study the nonlinear Hall effect
 and provide new perspectives for the application of this effect.
\end{abstract}

\maketitle

\section{I. Introduction}

In the past few decades, the study of topological phases has revealed that
both symmetry and topology have strong impact on the response of  materials
to external fields~\cite{Qi2008TFT}. One celebrated example is the anomalous Hall effect (AHE)
contributed by Berry curvature~\cite{Nagaosa2010AHE}. In the linear response regime, it is known that
the AHE identically vanishes in a time-reversal invariant system, owing to the constraint
of time-reversal symmetry on the distribution of Berry curvature~\cite{Xiao2010BP}. On the other hand,
the AHE will be robustly quantized in a two-dimensional time-reversal broken insulator if the
topological invariant, the net Chern number of its
occupied bands,  is nonzero~\cite{Haldane1988model,Yu2010QAHE,Chang2013QAHE}.
In physics, such robust quantized responses are appealing since their robustness allows interesting applications~\cite{Klitzing2019QHE},
meanwhile, they provide
an experimentally accessible and faithful way to extract the underlying  topological property of the concerned system.
However, quantized responses are commonly restricted to topological gapped systems and, being certain discrete quantities,
can only reflect the global topological property and its discrete change. As the local quantum geometry of the Bloch
wave functions, like
Berry curvature,
contains much richer information than its global counterpart, known as the topological invariant,
and it can have nontrivial distribution  even in a topologically trivial system,
non-quantized responses related to the local quantum geometry can appear in much more materials and can extract more
information about the band structure of the concerned system.

In recent years, it has been shown that several kinds of
nonlinear electromagnetic responses, including
injection current photogalvanic effect~\cite{Sipe2000,Hosur2011,deJuan2017},
shift current photogalvanic effect~\cite{Sipe2000,Tan2016,Cook2017,Yang2017shift,Yan2018shift,Osterhoudt2019}, nonlinear Hall effect~\cite{Moore2010BP,Sodemann2015BCD} and
so on, have a close connection to the local quantum geometry in momentum space~\cite{Morimoto2016,Ahn2020,Watanabe2021}.
Remarkably, the nonlinear Hall effect reveals that a Hall-like current can occur in time-reversal invariant
and inversion breaking systems as a second-order response to external electric fields~\cite{Moore2010BP,Sodemann2015BCD}. Being a second-order
intraband  effect, the Hall-like current is found to have a close connection with the dipole moment of the Berry curvature
in momentum space, the so-called Berry curvature dipole (BCD)~\cite{Sodemann2015BCD}. Mathematically, the BCD is an integral of the product of
local Berry curvature and velocity over the Fermi surface, so metals with prominent Berry curvature near the Fermi surface
are ideal platforms to observe this effect. Under this guiding principle, as band degeneracies are natural sources of divergent Berry curvature,
three-dimensional Weyl semimetals~\cite{Zhang2018BCDweyl,Rostami2018,Chen2019NHE,Matsyshyn2019NHE,Rostami2020,Gao2020NHE,Sinh2020NHE,Dzsaber2021Hall,Zeng2020BCD,Kumar2021NHE},
two-dimensional transition-metal dichalcogenides~\cite{Xu2018BCD,Ma2019NHE,Kang2019,Du2018NHE,You2018BCD,Zhang2018BCD,Wang2019NHE,Xiao2020BCD,Zhou2020NHE,Hu2020NHE,Huang2020NHE}, strained graphene~\cite{Battilomo2019BCD,Zhang2020NHE,Pantaleon2020BCD} and topological insulators close to the phase boundary~\cite{Facio2018}, which have either tilted
 gapless Weyl cones or tilted gapped Dirac cones, have been actively studied
both theoretically and experimentally\cite{Oritix2021review,Ma2021review}. As the nonlinear Hall effect is an effect related to Fermi surface,
it is noteworthy that doping is necessary for its observation in pristine gapped systems, such as topological insulators.

By far, most of works have focused on systems with sizable spin-orbit coupling, with  only a few exceptions~\cite{Battilomo2019BCD,Satyam2020BCD}.
When the spin-orbit coupling plays a important role in the band structure, the spin degrees of freedom
must be  taken into account and  the time-reversal symmetry operator satisfies $\mathcal{T}^{2}=-1$.
In the ten-fold way classification, it is known that the spinful time-reversal
symmetry allows the existence of strong topological insulators in two and three dimensions~\cite{Schnyder2008,kitaev2009periodic,Ryu2010}. In contrast,
when the spin-orbit coupling is negligible, the spin degrees of freedom can be neglected and
the time-reversal symmetry becomes its
spinless counterpart following $\mathcal{T}^{2}=1$. When only the spinless time-reversal
symmetry is present, it is known that the system belongs to the symmetry class AI in
the ten-fold way classification. It is worth noting that
the class AI does not support any strong topological insulator phase (``strong'' means that the gapless boundary states
are robust against symmetry-preserving perturbations and do not depend on which direction is chosen to be open) in one, two and
three dimensions~\cite{Schnyder2008,kitaev2009periodic,Ryu2010}, implying the absence of class AI strong topological insulators
in real materials. Because of this absence, materials belonging to the class AI
have attracted much less interest from the topological aspect. Accordingly,
the geometrical and topological properties of
the band structures as well as the related electromagnetic responses in this class of materials
have been poorly explored.

In this work, we consider both insulators and semimetals belonging to the class AI in two dimensions and
investigate, under different doping levels,  how the nonlinear Hall effect responds to the change of quantum geometry
and topology in band structures. For the class AI insulators in two dimensions,
Although there is  no strong topological phase,
we find that they can still be classified as topologically obstructed insulators and
trivial insulators if putting certain constraint on the Hamiltonians.
The remarkable distinctions between these two kinds of phases are manifested in whether there exist
boundary-direction-sensitive edge states and the distinct patterns of hybrid Wannier centers. Near the critical points between these two distinct phases,
we find that the BCDs near the band edge will become more prominent when the system gets closer
to the critical points. Notably, when going across the critical points,
the BCDs will dramatically reverse their signs.
For the semimetals, the topological properties and the locations of Dirac points (DPs) are of particular
interest. In this work, we consider the Mielke  model in which the DPs
can move along the Brillouin zone boundary and merge, accompanying with the change in quantum geometry and
topology~\cite{Mielke1991,Montambaux2018}. As the Berry curvature
is divergent at the DPs, we
find that the BCDs will sharply reverse their signs when the doping level crosses
the DPs. With the shift of locations of DPs in momentum and energy, the critical doping level
at which the BCDs sharply reverse their signs will accordingly change.
For both insulators and semimetals, the results suggest that the nonlinear Hall effect
can be quite remarkable in materials with negligible spin-orbit coupling and
is sensitive to the change in quantum geometry and topology.

The structure of the paper is as follows. In Sec. II,
we investigate the constraint put by the time-reversal symmetry and inversion symmetry on
the class AI models and briefly  review the nonlinear Hall effect.
In Sec. III, we consider two representative models realizing class AI insulators and study
the evolution of BCDs with respect to the change of band topology.
In Sec.IV, we take the Mielke model as a representative example to study
the evolution of nonlinear Hall effect with respect to the movement and mergence
of DPs.  We conclude with a detailed discussion in Sec.V.

\section{II. Theoretical formalism}

As the spinless time-reversal symmetry does not enforce Kramers degeneracy at any
momentum, and the nonlinear Hall effect is an effect related to Fermi surface,
it is a good starting point to consider two-band class AI Hamiltonians. It is known
that an arbitrary two-band Hamiltonian, in terms of the Pauli matrices,
can be expanded as
\begin{eqnarray}
\mathcal{H}(\bk)=\sum_{i=0,x,y,z}d_{i}(\bk)\sigma_{i}.\label{general}
\end{eqnarray}
Here we assume that the Pauli matrices $\sigma_{i}$ act on two orbital or sublattice degrees of freedom.
Without the loss of generality, we further assume that under the given
basis, the operator for the spinless time-reversal symmetry is simply
the complex conjugate operator $\mathcal{K}$, i.e. $\mathcal{T}=\mathcal{K}$.
Accordingly,  $\mathcal{T}^{2}=1$ is followed, and
the spinless time-reversal symmetry forces $\mathcal{H}^{*}(\bk)=\mathcal{H}(-\bk)$,
or equivalently,  $d_{0,x,z}(\bk)=d_{0,x,z}(-\bk)$, and $d_{y}(-\bk)=-d_{y}(\bk)$.

The energy spectra of the Hamiltonian in Eq.(\ref{general}) read
\begin{eqnarray}
E_{\pm}(\bk)=d_{0}(\bk)\pm\sqrt{d_{x}^{2}(\bk)+d_{y}^{2}(\bk)+d_{z}^{2}(\bk)}.
\end{eqnarray}
For the above spectra, it is easy to see that the existence of band degeneracy requires
the three components $\{d_{x}, d_{y}, d_{z}\}$ to vanish simultaneously at the same momentum.
When the dimension $\mathcal{D}\geq3$, stable band degeneracies can exist as the number of momentum variables
is equal to or larger than the number of constraint equations to fulfill.
In contrast, the band degeneracy can only be accidental in one and two dimensions if there is no additional
symmetry to force certain components of $\{d_{x}, d_{y}, d_{z}\}$ to vanish.

Besides the time-reversal symmetry, the inversion symmetry also puts
strong constraints on the Hamiltonian. When the Hamiltonian has inversion
symmetry, it needs to obey the constrain $\mathcal{P}\mathcal{H}(\bk)\mathcal{P}^{-1}=\mathcal{H}(-\bk)$
with $\mathcal{P}$ a unitary matrix satisfying $\mathcal{P}^{2}=1$.  As the time-reversal
operator and inversion operator commute, $\mathcal{P}$ has three possible choices
for the given Hamiltonian in Eq.(\ref{general}), i.e.
$\mathcal{P}=\sigma_{0}$, $\sigma_{x}$ or $\sigma_{z}$.  If $\mathcal{P}=\sigma_{0}$, then the
inversion symmetry forces $d_{0,x,y,z}(\bk)=d_{0,x,y,z}(-\bk)$. In combination
with the time-reversal symmetry, one immediately finds that $d_{y}(\bk)$ identically
vanishes throughout the Brillouin zone. If $\mathcal{P}=\sigma_{x} (\sigma_{z})$, then
$d_{0,x}(\bk)=d_{0,x}(-\bk)$ and $d_{y,z}(-\bk)=-d_{y,z}(\bk)$ ($d_{0,z}(\bk)=d_{0,z}(-\bk)$ and $d_{x,y}(-\bk)=-d_{x,y}(\bk)$),
which indicates  that $d_{z}(\bk)$ $(d_{x}(\bk))$ identically vanishes. In other words,
when both time-reversal symmetry and inversion symmetry are present, one of
the three components $\{d_{x}, d_{y}, d_{z}\}$ is forced to vanish identically,
consequently allowing the presence of stable point band degeneracies in two dimensions. A celebrated
example is the existence of Dirac points in graphene\cite{Castro2009RMP}.

For the two-band Hamiltonian in Eq.(\ref{general}),  the Berry curvature has a simple expression, which
reads~\cite{Qi2006spinhall}
\begin{eqnarray}
\Omega^{(\pm)}_{a}(\bk)=\pm\epsilon_{abc}\frac{\bd(\bk)\cdot(\partial_{b}\bd(\bk)\times \partial_{c}\bd(\bk))}{4d^{3}(\bk)}, \label{formula}
\end{eqnarray}
where the subscripts $\pm$ refer to the conduction and valence bands, respectively, $\epsilon_{abc}$ is the third-rank Levi-Civita symbol,  $\bd(\bk)=(d_{x}(\bk),d_{y}(\bk),d_{z}(\bk))$, $d(\bk)=|\bd(\bk)|$,
and $\partial_{a}\equiv\frac{\partial}{\partial k_{a}}$. In Eq.(\ref{formula}),
we have assumed the summation over the repeated indices. As is known,  the time-reversal symmetry
forces $\Omega^{(\pm)}_{a}(\bk)=-\Omega^{(\pm)}_{a}(-\bk)$ and the inversion symmetry forces
$\Omega^{(\pm)}_{a}(\bk)=\Omega^{(\pm)}_{a}(-\bk)$~\cite{Xiao2010BP}.
These two properties can also be simply inferred through the symmetry constraints on $d_{x,y,z}$ discussed previously.
Apparently, the coexistence of these two symmetries forces the Berry curvature to vanish
identically in momentum space as long as there is no band degeneracy at which the Berry
curvature is singular (note the denominator in Eq.(\ref{formula}) vanishes at the band degeneracy).
These symmetry constraints indicate that in a time-reversal invariant system, the inversion symmetry must be broken
to observe effects induced by Berry curvature.

It is known that a nonzero Berry curvature will contribute an anomalous velocity in
the semiclassic equations of motion~\cite{Xiao2010BP}, consequently, applying an electric field is possible
to generate a transverse Hall current even in the absence of an external magnetic field, which is
known as the anomalous Hall effect. In the linear response regime, the generated Hall current $j_{\alpha}$
and the applied electric field $\mathcal{E}_{b}$ are connected by the Hall conductivity $\sigma_{ab}$, i.e.
$j_{\alpha}=\sigma_{ab}\mathcal{E}_{b}$,
where $\sigma_{ab}$ is an antisymmetric tensor satisfying $\sigma_{ab}=-\sigma_{ba}$.
The general formula for $\sigma_{ab}$
takes the form (we take $\hbar=1$ throughout this paper)
\begin{eqnarray}
\sigma_{ab}=e^{2}\epsilon_{abc}\int\frac{d^{\mathcal{D}}k}{(2\pi)^{\mathcal{D}}}\sum_{\alpha}f^{(\alpha)}(\bk)\Omega_{c}^{(\alpha)}(\bk),\label{Hall}
\end{eqnarray}
where $\alpha$ runs over all bands and
$f^{(\alpha)}(\bk)=1/(1+\exp(E_{\alpha}(\bk)-\mu)/k_{B}T)$ is the Fermi-Dirac distribution function
for the $\alpha$th band,
with $\mu$ denoting the chemical potential, $k_{B}$ the Boltzmann constant and $T$ the temperature.
When the spinless time-reversal symmetry is conserved,
$E_{\alpha}(\bk)=E_{\alpha}(-\bk)$ and $\Omega_{c}^{(\alpha)}(-\bk)=-\Omega_{c}^{(\alpha)}(\bk)$.
Apparently, the integrand in Eq.(\ref{Hall}) is an odd function of momentum,
so $\sigma_{ab}$ is forced to vanish, indicating the absence of Hall current
in a time-reversal invariant system when restricting to the linear response regime.

Remarkably, the constraint on Hall current by the time-reversal symmetry can be lifted when
taking into account higher-order responses. Using the semiclassic equations of motion, Sodemann
and Fu derived that a dc as well as a second-harmonic Hall-like current  can appear in the
second-order response regime, as long as the inversion symmetry is broken~\cite{Sodemann2015BCD}. Under an oscillating electric field $\bE=\text{Re}\{\boldsymbol{\mathcal{E}}e^{i\omega t}\}$
with the amplitude vector $\boldsymbol{\mathcal{E}}$ and frequency $\omega$, the full Hall-like current can be written compactly as
$j_{a}=\text{Re}\{j_{a}^{(0)}+j_{a}^{(2)}e^{2i\omega t}\}$, where $j_{a}^{(0)}=\chi_{abc}\mathcal{E}_{b}\mathcal{E}_{c}^{*}$
describes the dc part, and $j_{a}^{(2)}=\chi_{abc}\mathcal{E}_{b}\mathcal{E}_{c}$ describes the second-harmonic part.
Interestingly, the coefficients in the two parts are equal and take the form~\cite{Sodemann2015BCD}
\begin{eqnarray}
\chi_{abc}=-\frac{e^{3}\tau}{2(1+i\omega \tau)}\epsilon_{adc}D_{bd}, \label{coeff}
\end{eqnarray}
where $\tau$ denotes the relaxation time which is assumed to be a constant,
and $D_{bd}$ is the BCD which takes the form
\begin{eqnarray}
D_{bd}&=&-\sum_{\alpha}\int\frac{d^{\mathcal{D}}k}{(2\pi)^{\mathcal{D}}}\partial_{b}f^{(\alpha)}(\bk)\Omega_{d}^{(\alpha)}(\bk)\nonumber\\
&=&\sum_{\alpha}\int\frac{d^{\mathcal{D}}k}{(2\pi)^{\mathcal{D}}}f^{(\alpha)}(\bk)\partial_{b}\Omega_{d}^{(\alpha)}(\bk).\label{berry}
\end{eqnarray}
The first line contains a factor $\partial_{b}f^{(\alpha)}$, which is equal to $-\partial_{b}E_{\alpha}\delta(\mu-E_{\alpha})$
at the zero-temperature limit. The formula indicates that the BCD is an integral of the product of Berry curvature
and Fermi velocity over the Fermi surface.

It is worth noting that the Berry curvature has only one component in two dimensions, namely $\Omega_{z}(\bk)$. As the second
label of $D_{bd}$ is fixed in this dimension, we will follow Ref.~\cite{Sodemann2015BCD} and
adopt the shorthand notation $D_{x}$ and $D_{y}$
to substitute $D_{xz}$ and $D_{yz}$ for a simplification of the notation.

\section{III. Nonlinear Hall effect in doped class AI insulators}
\subsection{A. Insulators with low-energy linear Dirac cones}

\begin{figure*}[htbp]
	\begin{center}
		\includegraphics[width=0.6\textwidth]{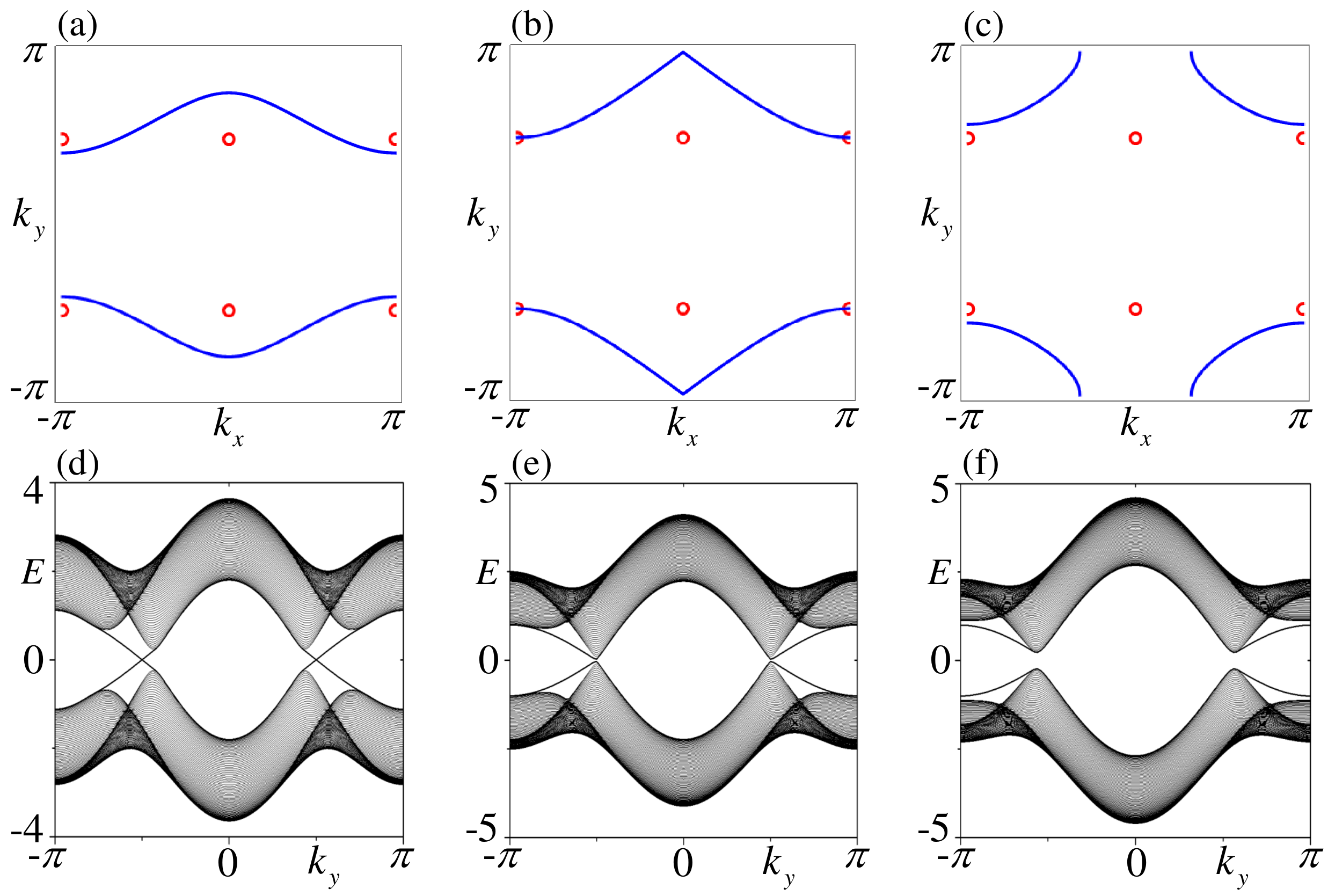}
		\caption{(Color online) (a)-(c) Configurations of BIS (blue lines) and DPs (red points) in the bulk Brillouin zone. (d)-(e)
Energy spectra for a cylinder geometry with open boundary conditions in the $x$ direction and
periodic boundary conditions in the $y$ direction. Common parameters are $t_{0}=0$, $t_{1}=t_{3}=1$ and $t_{2}=t_{4}=2$.
$m=0.5$ in (a)(d), $m=1$ in (b)(e), and $m=1.5$ in (c)(f). }
		\label{bulkedge}
	\end{center}
\end{figure*}

Now let us focus on concrete class AI models  and explore their geometrical and topological properties, as well as
the nonlinear Hall effect. As the first representative example, we consider that the four components of $d_{i}$ take
the following form,
\begin{eqnarray}
d_{0}(\bk)&=&t_{0}\sin k_{x}\sin k_{y}, \nonumber\\
d_{x}(\bk)&=&t_{1}\cos k_{y},\quad
d_{y}(\bk)=t_{2}\sin k_{x},\nonumber\\
d_{z}(\bk)&=&(m+t_{3}\cos k_{x}+t_{4}\cos k_{y}).\label{model1}
\end{eqnarray}
For notational simplicity, the lattice constants are set to unity
throughout this paper. Through dimensional analysis, it is easy to
find that the BCD is of the dimension of length in two dimensions, so
the full expression  for BCD can be simply restored by  multiplying the dimensionless
BCD with the corresponding lattice constant in real materials.
As $d_{0,x,z}(\bk)=d_{0,x,z}(-\bk)$, and $d_{y}(-\bk)=-d_{y}(\bk)$, the Hamiltonian
has the spinless time-reversal symmetry, but the inversion symmetry is broken according to
our previous analysis.
When $t_{0}=0$, the Hamiltonian
has one additional mirror symmetry, i.e. $\mathcal{M}_{y}\mathcal{H}(k_{x},k_{y})\mathcal{M}_{y}^{-1}
=\mathcal{H}(k_{x},-k_{y})$ with $\mathcal{M}_{y}=\sigma_{0}$. The mirror symmetry
$\mathcal{M}_{y}$ will force the Berry curvature
to be odd in $k_{y}$, i.e. $\Omega_{z}^{(\pm)}(k_{x},k_{y})=-\Omega_{z}^{(\pm)}(k_{x},-k_{y})$. As the time-reversal symmetry
forces $\Omega_{z}^{(\pm)}(-k_{x},-k_{y})=-\Omega_{z}^{(\pm)}(k_{x},k_{y})$, a combination leads to
$\Omega_{z}^{(\pm)}(k_{x},k_{y})=\Omega_{z}^{(\pm)}(-k_{x},k_{y})$. Because
the Fermi distribution function $f^{(\pm)}(k_{x},k_{y})$ is also an even function
of $k_{x}$ when $t_{0}=0$,  one can easily infer from Eq.(\ref{berry}) that
$D_{x}$ is identically equal to zero for this case. According to Eq.(\ref{coeff}),
$D_{x}=0$ implies $\chi_{yxx}=\chi_{xxy}=0$. For linearly polarized light,
$\chi_{yxx}=0$ implies that the Hall-like current is absent when the electric
vector is in the $x$ direction. This result can also be intuitively understood by
noting that under the mirror reflection $\mathcal{M}_{y}$, the current and electric field
follow the change: $\{j_{x},j_{y}; \mathcal{E}_{x}, \mathcal{E}_{y}\}\rightarrow \{j_{x},-j_{y}; \mathcal{E}_{x}, -\mathcal{E}_{y}\}$.
Accordingly, the existence of mirror symmetry $\mathcal{M}_{y}$ implies
$j_{y}=\chi_{yxx}\mathcal{E}_{x}\mathcal{E}_{x}=-j_{y}$ and $j_{x}=\chi_{xxy}\mathcal{E}_{x}\mathcal{E}_{y}=-\chi_{xxy}\mathcal{E}_{x}\mathcal{E}_{y}$,
so $\chi_{yxx}=\chi_{xxy}=0$. When $t_{0}\neq0$, because the $d_{0}$ term does not affect
the Berry curvature, the relation $\Omega_{z}^{(\pm)}(k_{x},k_{y})=\Omega_{z}^{(\pm)}(-k_{x},k_{y})$
remains hold. However, the Fermi distribution function $f^{(\pm)}(k_{x},k_{y})$ is no longer invariant under
$k_{x}\rightarrow -k_{x}$,  so a finite $D_{x}$ becomes possible.

Before calculating $D_{x}$ and $D_{y}$, we first explore the
geometrical and topological properties of the Hamiltonian.
As none of the three components $\{d_{x},d_{y},d_{z}\}$ vanishes identically, the Hamiltonian in
general has a gapped spectra and thus describes an insulator.
Despite that the Hamiltonian, according
to the ten-fold way classification, cannot realize strong topological phases whose gapless boundary states
are insensitive to the orientation of boundary, it can be still topologically nontrivial
in a general sense. That is, in some regimes,
the Hamiltonian, if putting certain constraint,
cannot be adiabatically deformed to the trivial atomic limit
without the closing of bulk gap. In such obstructed regimes,
the Hamiltonian  can still harbor gapless edge states on some boundary
and meanwhile the hybrid Wannier centers of its bands also display
nontrivial features.

We first provide an intuitive bulk picture for the existence of
obstructed regime. For the convenience of discussion, we term
the contour satisfying $d_{z}(\bk)=0$ as band inversion surface (BIS),
and the points simultaneously satisfying $d_{x}(\bk)=d_{y}(\bk)=0$
as Dirac points (DPs). As long as the BIS can be adiabatically deformed to vanish
without closing the bulk gap, the phase is adiabatically
connected to the
atomic limit ($m\rightarrow\infty$) and is thus topologically trivial, otherwise the system
falls into the obstructed regime.

The quantum anomalous Hall insulator with inversion symmetry can
be taken as the simplest realization of topologically obstructed phase. Let us take
the Qi-Wu-Zhang model for illustration, where $d_{x}=\sin k_{y}$, $d_{y}=-\sin k_{x}$,
and $d_{z}(\bk)=c(2-\cos k_{x}-\cos k_{y}-e_{s})$~\cite{Qi2006spinhall}. For this inversion symmetric model,
the DPs satisfying $d_{x}(\bk)=d_{y}(\bk)=0$ are pinned at the four time-reversal invariant
momenta. When the BIS satisfying $d_{z}(\bk)=0$ encloses one of the DPs, i.e. $0<|e_{s}|<2$ with $c\neq0$,
it is easy to see that, because of the periodicity of Brillouin zone,  the BIS
cannot be adiabatically deformed to vanish without crossing the DPs (the bulk
gap gets closed when they cross), so such a situation corresponds to the obstructed regime.
For the Qi-Wu-Zhang model, it is apparent that the four unmovable DPs at the time-reversal invariant momenta are essential
for the obstruction.

For the Hamiltonian in Eq.(\ref{model1}),
it is readily found that there are also four DPs. However, they
are not located at the time-reversal invariant momenta. Instead,
they are located at $(k_{x},k_{y})=(0,\pm\pi/2)$ and $(\pi,\pm\pi/2)$ in
the first Brillouin zone. As these four DPs are not pinned at the
four special time-reversal invariant momenta, they are in principle
movable and can annihilate. For instance, one can add a symmetry-preserving term
of the form $\delta\sigma_{x}$  to the Hamiltonian in Eq.(\ref{model1}),
then the DPs will annihilate and disappear when $|\delta|>|t_{1}|$.
The absence of unremovable DPs can be taken as the underlying reason why the class AI
does not host strong topological phase in two dimensions. For real materials,
however, if the DPs are there, their positions may be quite stable under the change of
certain experimental conditions, such as the pressure.
With this in mind, here we will make an assumption
that the DPs, though their locations are not at the time-reversal invariant momenta,  are fixed in momentum space,
and only the BIS changes. Under this assumption, the class AI insulators can also
be classified as topologically obstructed insulators and trivial insulators according
to the configurations of BIS and DPs.

Without the loss of generality, in the following
we take $t_{1}=t_{3}=1$ and $t_{2}=t_{4}=2$ for a concrete discussion.
For this set of parameters, $-1<m<1$ is found to be the obstructed regime
of the Hamiltonian in Eq.(\ref{model1}).
To see this in an intuitive way, we show the BIS and DPs together in the first
Brillouin zone.
The result for $m=0.5$ is shown in Fig.\ref{bulkedge}(a).
It is readily found that under this parameter condition,
the BIS (blue lines) cannot be continuously deformed to vanish without crossing the fixed
DPs (red points). As the bulk gap gets closed when the BIS and DPs cross, it means that
if starting with $m=0.5$, the BIS cannot be adiabatically  deformed to vanish without
closing the bulk gap, so it corresponds to the obstructed regime.
At $m=1$, the crossing of BIS and DPs shown in Fig.\ref{bulkedge}(b) indicates that the
bulk gap is closed at this critical point. The result for $m=1.5$ is shown
in Fig.\ref{bulkedge}(c). It is readily found that under this parameter condition,
the BIS can be continuously deformed to vanish without closing the bulk gap,
indicating that the Hamiltonian now falls into the trivial regime.

\begin{figure}
\subfigure{\includegraphics[width=9cm, height=8cm]{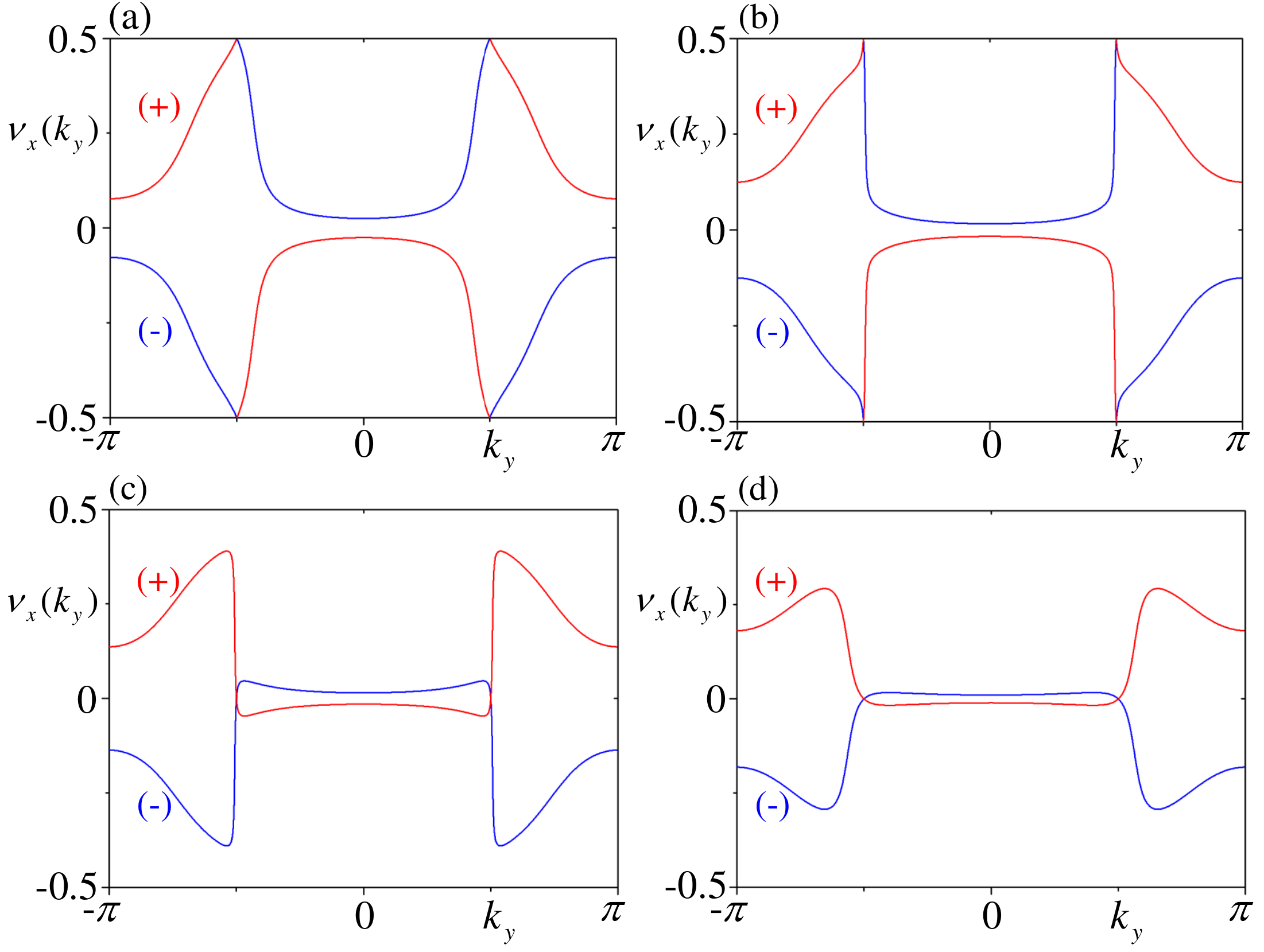}}
\caption{ (Color online) The hybrid Wannier centers $\nu_{x}^{\pm}(k_{y})$. (a) $m=0.5$, (b) $m=0.95$, (c)
$m=1.05$, (d) $m=1.5$. The hybrid Wannier centers undergo a dramatic change in structure when $m$ crosses
the critical point at $m=1$. Common parameters are $t_{0}=0$, $t_{1}=t_{3}=1$ and $t_{2}=t_{4}=2$.
}  \label{wannier}
\end{figure}

Now we consider a cylinder geometry with open boundary conditions in the $x$ direction and
periodic boundary conditions in the $y$ direction and show the corresponding energy spectra under the
same parameter conditions as in Figs.\ref{bulkedge}(a)-(c). According to the results in Fig.\ref{bulkedge}(d)-(f), it is
readily found that the Hamiltonian harbors gapless edge states only in the obstructed regime.
It is worth noting that if the boundary conditions are reversed, namely open boundary conditions in the $y$ direction and
periodic boundary conditions in the $x$ direction, then gapless edge states are always absent
(so not shown explicitly), regardless of whether the Hamiltonian is in the obstructed regime or not.
This is consistent with the fact that the Hamiltonian cannot realize a strong topological phase for which
the existence of gapless edge states does not depend on the choice of boundary.

\begin{figure*}[htbp]
	\begin{center}
		\includegraphics[width=0.9\textwidth]{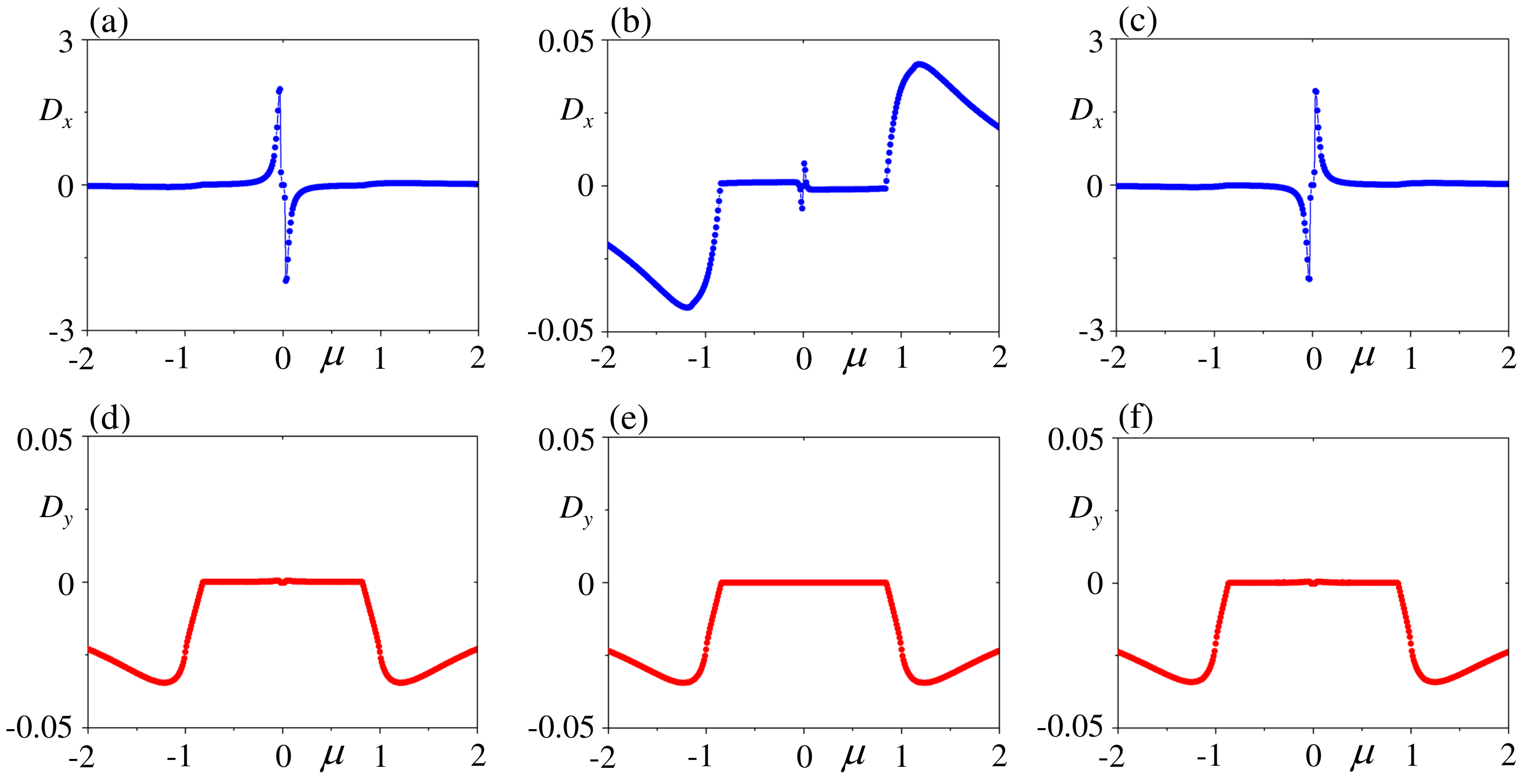}
		\caption{(Color online) The evolution of $D_{x}$ and $D_{y}$ across the critical point at $m=1$. Common parameters
are $t_{0}=0$, $t_{1}=t_{3}=1$ and $t_{2}=t_{4}=2$. $m=0.95$ in (a)(d), $m=1$ in (b)(e), and $m=1.05$ in (c)(f). }
		\label{nlhe}
	\end{center}
\end{figure*}

In the following, we further show the hybrid Wannier centers which
are determined by calculating the corresponding Wilson loop.
For the two-band Hamiltonian in Eq.(\ref{model1}), the Wilson loops in the
$x$ and $y$ directions have the form~\cite{Vanderbilt2018book}
\begin{eqnarray}
W^{(\pm)}_{x}(k_{y})&=&\prod_{n=0}^{N_{x}-1}\langle u^{(\pm)}(\bk_{n,x})|u^{(\pm)}(\bk_{n+1,x})\rangle,\nonumber\\
W_{y}^{(\pm)}(k_{x})&=&\prod_{n=0}^{N_{y}-1}\langle u^{(\pm)}(\bk_{n,y})|u^{(\pm)}(\bk_{n+1,y})\rangle, \nonumber
\end{eqnarray}
where $\bk_{n,x}=\bk+n\Delta \bk_{x}$ with $\Delta \bk_{x}=(2\pi/N_{x},0)$ and
$\bk_{n,y}=\bk+n\Delta \bk_{y}$ with $\Delta \bk_{y}=(0,2\pi/N_{y})$, $|u^{\pm}(\bk)\rangle$ denote
the wave functions of the upper and lower bands, respectively, and $|u^{\pm}(\bk)\rangle$
obey the periodic gauge, i.e. $|u^{\pm}(\bk)\rangle=|u^{\pm}(\bk+\bG)\rangle$ with $\bG$ the reciprocal lattice vector.
Based on the Wilson loops, the hybrid Wannier centers are given by~\cite{Vanderbilt2018book}
\begin{eqnarray}
\nu_{x}^{(\pm)}(k_{y})&=&-\frac{1}{2\pi}\text{Im}\ln W_{x}(k_{y}), \nonumber\\
\nu_{y}^{(\pm)}(k_{x})&=&-\frac{1}{2\pi}\text{Im}\ln W_{y}(k_{x}),
\end{eqnarray}
where $\text{Im}$ denotes to take the imaginary part. As $\nu_{x}^{(\pm)}$ and $v_{y}^{(\pm)}$
are gauge-invariant modulo $1$, in this work we choose the domain $\nu_{x,y}^{(\pm)}\in(-0.5,0.5]$.
Because the $d_{0}(\bk)$ term does not affect the wave functions, we find that the mirror
symmetry $\mathcal{M}_{y}$ forces $\nu_{y}^{(\pm)}(k_{x})$ to take the trivial zero value
for arbitrary $k_{x}$, which is consistent with the absence of gapless excitations
on the $y$-normal edges. The numerical results for $\nu_{x}^{(\pm)}(k_{y})$
are shown in Figs.\ref{wannier}(a)-(d). It is readily seen that $\nu_{x}^{(\pm)}(k_{y})$
display remarkable difference in the obstructed regime and trivial regime. In
the obstructed regime, $\nu_{x}^{(\pm)}(k_{y})$ will always cross the domain boundary $\nu=0.5$ (this value corresponds
to a $\pi$ Berry phase as the Berry phase $\phi=2\pi\nu$)
at $k_{y}=\pm\pi/2$. In contrast, in the trivial regime, $\nu_{x}^{(\pm)}(k_{y})$ cross
the domain center $\nu=0$ at $k_{y}=\pm\pi/2$.  Here the reason for the crossings
to be fixed at $k_{y}=\pm\pi/2$ is simply because the DPs satisfying
$d_{x}(\bk)=d_{y}(\bk)=0$ are fixed
at $(k_{x},k_{y})=(0,\pm\pi/2)$ and $(\pi,\pm\pi/2)$.

As the Hamiltonian has a dramatic change in geometrical and topological properties
when going from the obstructed regime to the trivial regime,
it is interesting to investigate how the nonlinear Hall effect responds to this change.
To make $D_{x}$ nonzero, we consider $t_{0}$ to be finite. Besides the breaking of
mirror symmetry, here another effect of the $t_{0}$ term is tilting the
Dirac cones which describes the low-energy physics. To see this, we consider $m\rightarrow1$,
so that the low-energy physics is described by the Dirac Hamiltonians at $\bK_{\pm}=(\pi,\pm\pi/2)$.
By expanding  the Hamiltonian in Eq.(\ref{model1}) around $\bK_{\pm}$ to the linear order in momentum, the
Dirac Hamiltonians have the form
\begin{eqnarray}
H_{\bK_{\pm}}(\bq)=\mp t_{0}q_{x}\mp t_{1}q_{y}\sigma_{x}-t_{2}q_{x}\sigma_{y}+\tilde{m}\sigma_{z}, \label{Dirac}
\end{eqnarray}
where $\tilde{m}=m-t_{3}$ and $\bq$ represents the momentum measured from $\bK_{+}$ or $\bK_{-}$. It becomes
apparent that the $t_{0}$ term tilts the two
Dirac cones along the $x$ direction in an opposite way. Restricting to the low-energy Hamiltonians, the Berry curvatures
around $\bK_{\pm}$ have the form
\begin{eqnarray}
\Omega_{\bK_{+}}^{(\pm)}(\bq)=\mp\frac{t_{1}t_{2}\tilde{m}}{2(t_{1}^{2}q_{y}^{2}+t_{2}^{2}q_{x}^{2}+\tilde{m}^{2})^{3/2}},\nonumber\\
\Omega_{\bK_{-}}^{(\pm)}(\bq)=\pm\frac{t_{1}t_{2}\tilde{m}}{2(t_{1}^{2}q_{y}^{2}+t_{2}^{2}q_{x}^{2}+\tilde{m}^{2})^{3/2}}.\label{valley}
\end{eqnarray}
One can find the Berry curvatures are even under $q_{y}\rightarrow-q_{y}$.
Because the Fermi-Dirac distribution function is also even under $q_{y}\rightarrow-q_{y}$,
it can be readily inferred from Eq.(\ref{berry}) that $D_{y}$ will vanish if restricting
to the above linear Dirac Hamiltonians. This analysis suggests that if $D_{y}$ is nonzero
for the full Hamiltonian, the nonzero parts come from higher-order contributions and
thus should be small when the system is close to the critical point and the chemical potential is located near the band edge.

Depending on the extent of tilt, the Dirac cones are commonly classified as type-I and
type-II ones. For this model, as the tilt only affects the result quantitatively,
we will focus on type-I Dirac cones, i.e. $|t_{0}|<|t_{2}|$, so that the Hamiltonian
describes a true insulator when the chemical potential is located within the gap of
 Dirac cones. In Fig.\ref{nlhe}, we show the evolution of $D_{x}$ and $D_{y}$
across the critical point at $m=1$ for a broad range of doping levels. From
Figs.\ref{nlhe}(a)-(c), it is readily seen that $D_{x}$ dramatically changes its sign when
the system crosses the critical point. Furthermore, on both sides, $D_{x}$ takes a very large
value when the chemical potential is located near the band edge. In contrast, from
Figs.\ref{nlhe}(d)-(f), one can find that $D_{y}$ does not display
any dramatic change and is vanishingly small when the doping level is lower than a threshold, agreeing with the previous analysis
based on the low-energy linear Dirac Hamiltonians.

The dramatic sign change of $D_{x}$ across the critical point can be understood from Eq.(\ref{valley}).
Since the Dirac mass $\tilde{m}$ changes sign when the system crosses the critical point, one can
infer from Eq.(\ref{valley}) that the Berry curvatures at $\bK_{\pm}$ switch their sign.
As the locations of the two Dirac cones are fixed in momentum space, the sign change
in Berry curvatures directly leads to the sign change of $D_{x}$. On the other hand,
the large value of $D_{x}$ near the band edge is a consequence of the large Berry curvature, the
large density of states and the tilt which makes the Fermi velocity finite
at the band edge.

\subsection{B. Insulators with low-energy semi-Dirac cones}

For the first model, the low-energy physics near the phase boundary is
captured by linear Dirac cones.
In this section, we replace the linear Dirac cones by semi-Dirac cones whose
dispersions are quadratic in one direction and linear in other directions
when the Dirac mass vanishes. To realize such semi-Dirac cones, we substitute the DPs by semi-Dirac points (SDP) which correspond to the
mergence of two Dirac points with opposite winding numbers. To be specific,
we consider that $d_{i}$ have the form
\begin{eqnarray}
d_{0}(\bk)&=&0, \quad d_{x}(\bk)=t_{1}\sin k_{x}\sin k_{y}, \nonumber\\
d_{y}(\bk)&=&t_{2}\sin k_{x}+t_{3}\sin k_{y}, \nonumber\\
d_{z}(\bk)&=&(m+t_{4}\cos k_{x}+t_{5}\cos k_{y}).\label{model2}
\end{eqnarray}
Here we set $d_{0}(\bk)=0$ because the Hamiltonian
does not have any crystalline symmetry even when the $d_{0}$ term is absent.
In addition, here as the SDPs satisfying $d_{x}(\bk)=d_{y}(\bk)=0$ are fixed at time-reversal invariant momentum,
one cannot add a term to tilt the linear direction while preserving the
time-reversal symmetry. Although the SDPs
correspond to a critical situation and they can be gapped by an arbitrary small time-reversal symmetric
perturbation of the form $\delta\sigma_{x}$, one can view two nearby DPs with opposite winding numbers
as an effective SDP and the above Hamiltonian can be applied to describe
such more realistic situations.

In the first model, we have shown that the relative configuration between BIS and DPs
determines whether the system falls into the obstructed regime or the trivial regime.
Without the loss of generality,  we take $t_{1,2,3,4,5}=1$
for a concrete discussion in the following. For this set of parameters, the BIS encloses the SDP
at the $\boldsymbol{\Gamma}=(0,0)$ point when $-2<m<0$ and encloses the SDP
at the $\boldsymbol{M}=(\pi,\pi)$ point when $0<m<2$. At $m=\pm2$ and $m=0$,
the BIS crosses one and two SDPs, respectively, leading to the closing of bulk gap.
According to the analysis,  if we make a similar assumption as the previous case
that the SDPs are fixed there, then
$0<|m|<2$ corresponds to the obstructed regime and $|m|>2$ corresponds to
the trivial regime.

\begin{figure}
\subfigure{\includegraphics[width=9cm, height=7cm]{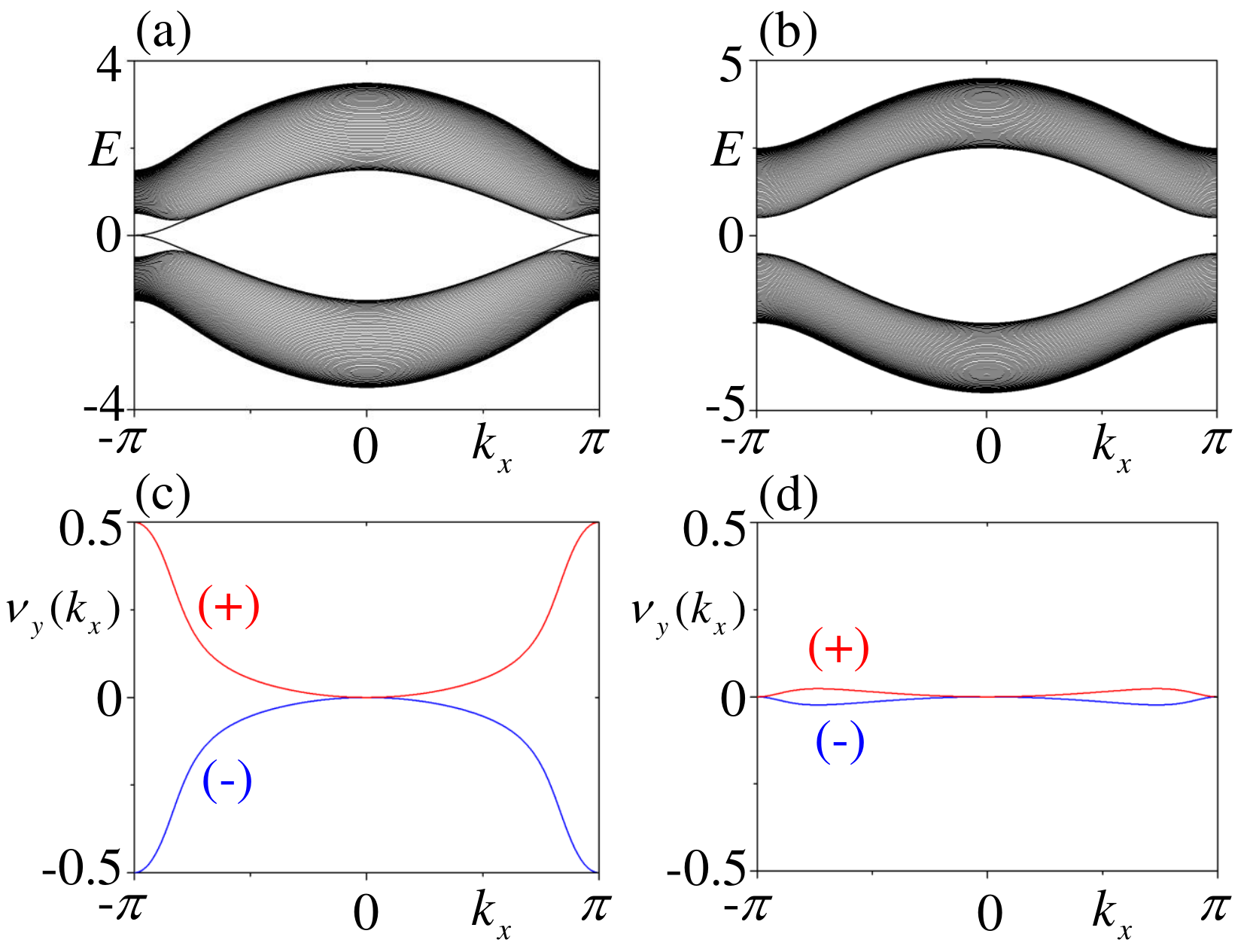}}
\caption{ (Color online) (a)(b) show the energy spectra under a cylinder geometry with open boundary conditions
in the $y$ direction and periodic boundary conditions in the $x$ direction. (c)(d) show
the hybrid Wannier centers $\nu_{y}^{(\pm)}(k_{x})$. Common parameters are $t_{1}=t_{2}=t_{3}=t_{4}=t_{5}=1$.
$m=1.5$ in (a)(c), and $m=2.5$ in (b)(d).
}  \label{edgewannier}
\end{figure}

In Fig.\ref{edgewannier}, the energy spectra for a cylinder geometry as well as the hybrid Wannier
centers are shown. Because the Hamiltonian does not change under the exchange $k_{x}\rightarrow k_{y}$
for the chosen set of parameters, the results shown in Fig.\ref{edgewannier} does not change
if $k_{x}$ is substituted by $k_{y}$, so here we only show one of them to avoid repetition.
The results for $m=1.5$ are shown in Figs.\ref{edgewannier}(a)(c). One can find that the open edges
harbor mid-gap states, and the hybrid Wannier centers cross the domain boundary $\nu=0.5$ at
$k_{x}=\pi$. In contrast,
when $m=2.5$, which falls into the trivial regime, one can find from Figs.\ref{edgewannier}(b)(d)
that the open edges do not harbor any mid-gap state, and
the hybrid Wannier centers instead only cross the domain center $\nu=0$ at both $k_{x}=0$ and $k_{x}=\pi$.

\begin{figure*}[htbp]
	\begin{center}
		\includegraphics[width=0.9\textwidth]{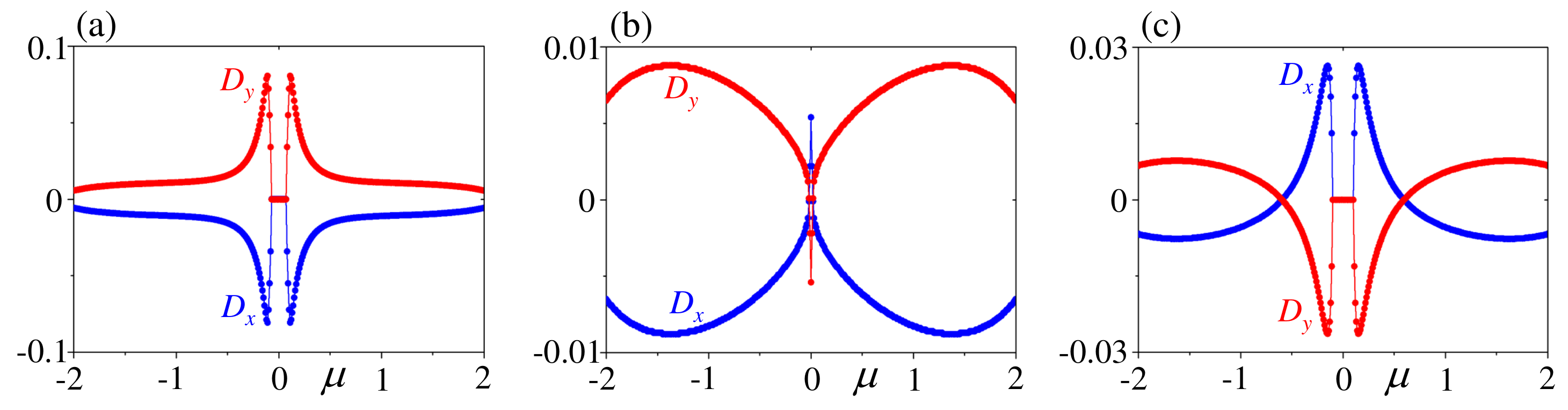}
		\caption{(Color online) The evolution of $D_{x}$ and $D_{y}$ across the critical point at $m=2$. Common parameters
are $t_{1}=t_{2}=t_{3}=t_{4}=t_{5}=1$. $m=1.9$ in (a), $m=2$ in (b), and $m=2.1$ in (c). $D_{x}=-D_{y}$ in (a)-(c).}
		\label{bcp2}
	\end{center}
\end{figure*}

Now we investigate the evolution of $D_{x}$ and $D_{y}$ across the critical points. Without
the loss of generality, we focus on the critical point at $m=2$. Near this critical point,
the band edge is located at $\boldsymbol{M}=(\pi,\pi)$, so we do a low-energy expansion around
this point. For simplicity, we only keep the leading-order term for each of the $d_{i}$ components.
As a result, the low-energy Hamiltonian reads
\begin{eqnarray}
H_{\boldsymbol{M}}(\bq)=t_{1}q_{x}q_{y}\sigma_{x}-(t_{2}q_{x}+t_{3}q_{y})\sigma_{y}+\tilde{m}\sigma_{z}.
\end{eqnarray}
Here $\tilde{m}=m-t_{4}-t_{5}$ and $\bq$ represents the momentum measured from $\boldsymbol{M}$.
At $\tilde{m}=0$, the dispersion of this Hamiltonian is quadratic in the direction
satisfying $t_{2}q_{x}+t_{3}q_{y}=0$. The Berry curvatures for this low-energy Hamiltonian
read
\begin{eqnarray}
\Omega_{\boldsymbol{M}}^{(\pm)}(\bq)=\pm\frac{t_{1}(t_{2}q_{x}-t_{3}q_{y})\tilde{m}}
{2\left[t_{1}^{2}q_{x}^{2}q_{y}^{2}+(t_{2}q_{x}+t_{3}q_{y})^{2}+\tilde{m}^{2}\right]^{3/2}}.\label{berryM}
\end{eqnarray}
The Berry curvatures have two interesting properties. One is still the sign change when the
system crosses the critical point. The second one is that the Berry curvature is odd under the
exchange $q_{x}\leftrightarrow q_{y}$ when $t_{2}=t_{3}$. The second property can be simply inferred from the
definition of Berry curvature which is antisymmetric about the two orthogonal momenta in two dimensions (see
Eq.(\ref{formula})).

The evolutions of $D_{x}$ and $D_{y}$ crossing the critical point $m=2$ are shown in Fig.\ref{bcp2}.
The results indicate that $D_{x}$ and $D_{y}$ will dramatically change their signs when
the system crosses the critical point. In addition, one can find that $D_{x}=-D_{y}$ is always hold for
the given parameters,  regardless of the doping level. These two main features can be explained by
the two properties of Berry curvatures discussed above.

Compared to Fig.\ref{nlhe}, one can find that when the doping level is near the band edge,
$D_{x}$ in Fig.\ref{bcp2} is much smaller. This remarkable difference can be understood through
the difference in their Berry curvatures. According to Eq.(\ref{valley}), the Berry curvatures for
the linear Dirac cones are peaked
 at $\bq=0$ with a height proportional to $1/2\tilde{m}^{2}$. In contrast, according to Eq.(\ref{berryM}),
the Berry curvature for the semi-Dirac cone is linear in momentum~\cite{Satyam2020BCD},
and thus vanish in the limit $\bq\rightarrow0$ when $\tilde{m}$
is still finite.

\begin{figure}
\subfigure{\includegraphics[width=9cm, height=5cm]{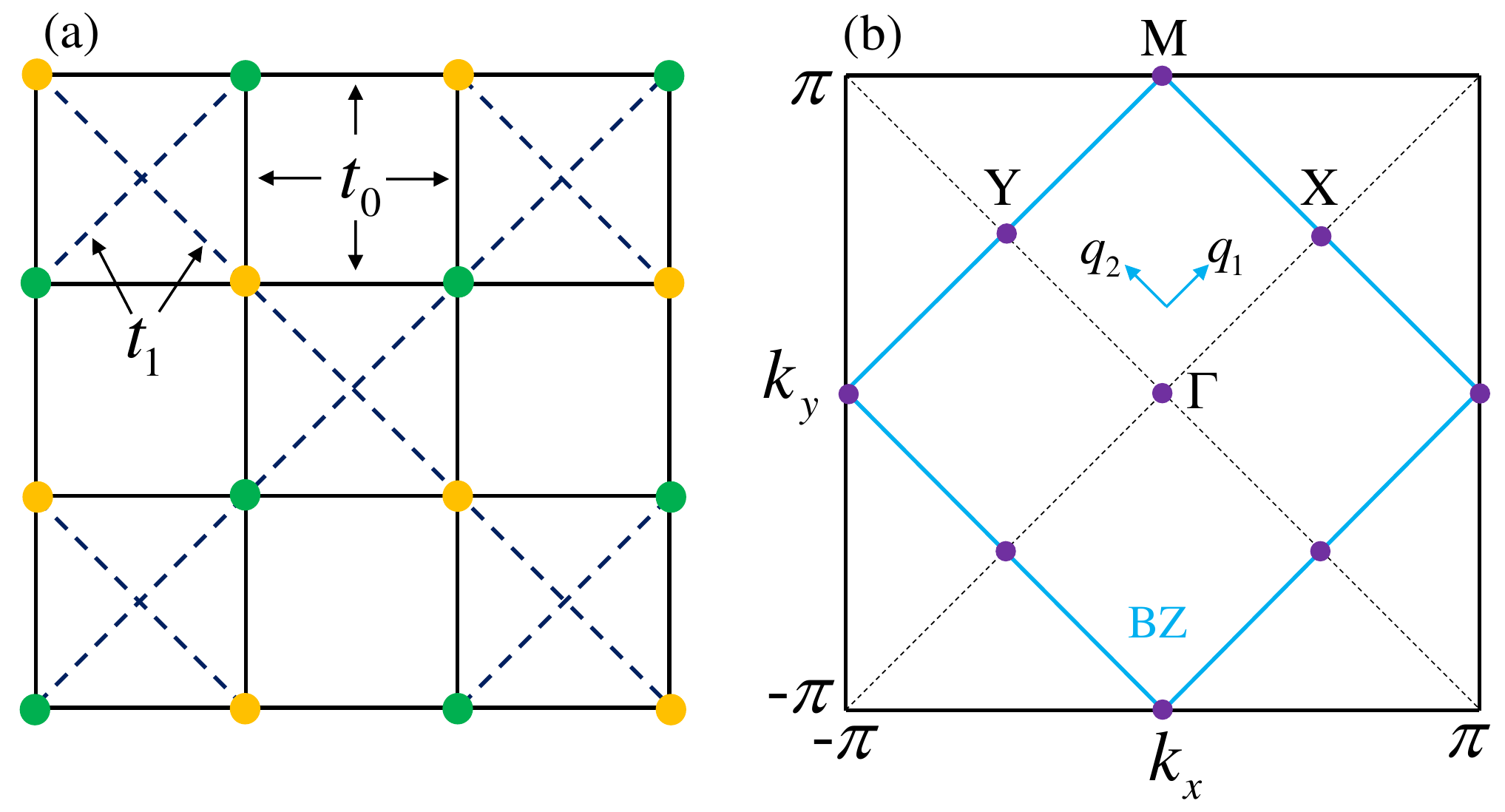}}
\caption{ (Color online) (a) A schematic diagram of the lattice and hoppings. $t_{0}$ denotes the nearest neighbor
hopping, and $t_{1}$ denotes the next-nearest neighbor hopping. The yellow and green dots denote two
kinds of sublattice. (b) The cyan square is the first Brillouin zone of the lattice in (a).
}  \label{sketch}
\end{figure}

\subsection{IV. Nonlinear Hall effect in doped class AI semimetals}

By far the two studied models have gapped band structure and the locations
of linear or semi-Dirac cones are fixed.
As semimetals are also of great interest, below we consider a simple
two-band model which has gapless band structure in a broad
regime. To be specific, we consider the staggered Mielke model whose real-space
hoppings and first Brillouin zone are illustrated in Figs.\ref{sketch}(a) and (b), respectively.
For this model, the four components of $d_{i}$ take the form~\cite{Mielke1991,Montambaux2018}
\begin{eqnarray}
d_{0}(\bk)&=&2t_{1}\cos k_{x}\cos k_{y},\nonumber\\
d_{x}(\bk)&=&2t_{0}(\cos k_{x}+\cos k_{y})\cos k_{y},\nonumber\\
d_{y}(\bk)&=&2t_{0}(\cos k_{x}+\cos k_{y})\sin k_{y},\nonumber\\
d_{z}(\bk)&=&2t_{1}\sin k_{x}\sin k_{y}+\delta.
\end{eqnarray}
This Hamiltonian has quite a few salient features. First, it is the simplest
model which can realize a flat band with nontrivial geometrical and topological properties. Second, while this Hamiltonian does not have
inversion symmetry, the existence of a glide symmetry allows the band structure to have stable band degeneracies at the
Brillouin zone boundary  when the on-site offset potential
$\delta$ is smaller than twice the diagonal hopping $t_{1}$, i.e. $|\delta|<2|t_{1}|$.
Third, with the variation of $\delta$, the topological properties of the band degeneracies will change,
and the band degeneracies will move in momentum space,  merge when $|\delta|=2|t_{1}|$, and
then annihilate and disappear when $|\delta|>2|t_{1}|$.

\begin{figure*}[htbp]
	\begin{center}
		\includegraphics[width=0.9\textwidth]{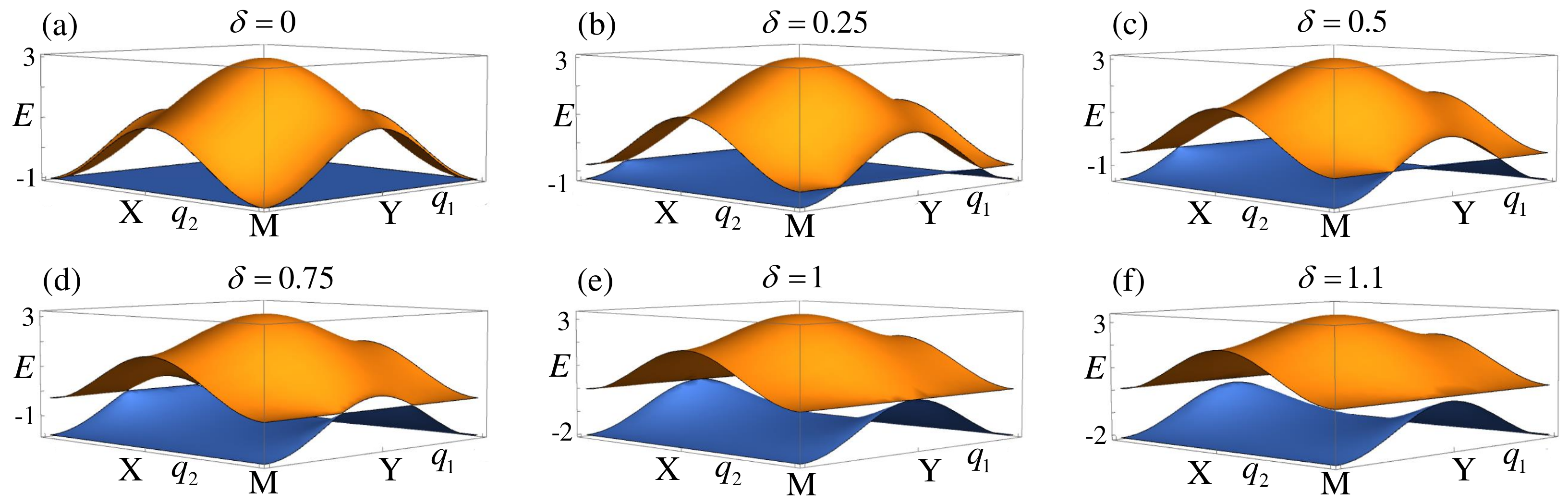}
		\caption{(Color online) Bulk energy spectra of the Mielke model. Common parameters are $t_{0}=t_{1}=0.5$. (a)
$\delta=0$, (b) $\delta=0.25$, (c) $\delta=0.5$, (d) $\delta=0.75$, (e) $\delta=1$ and (f) $\delta=1.1$.  }
		\label{spectrum}
	\end{center}
\end{figure*}

\begin{figure*}[htbp]
	\begin{center}
		\includegraphics[width=0.9\textwidth]{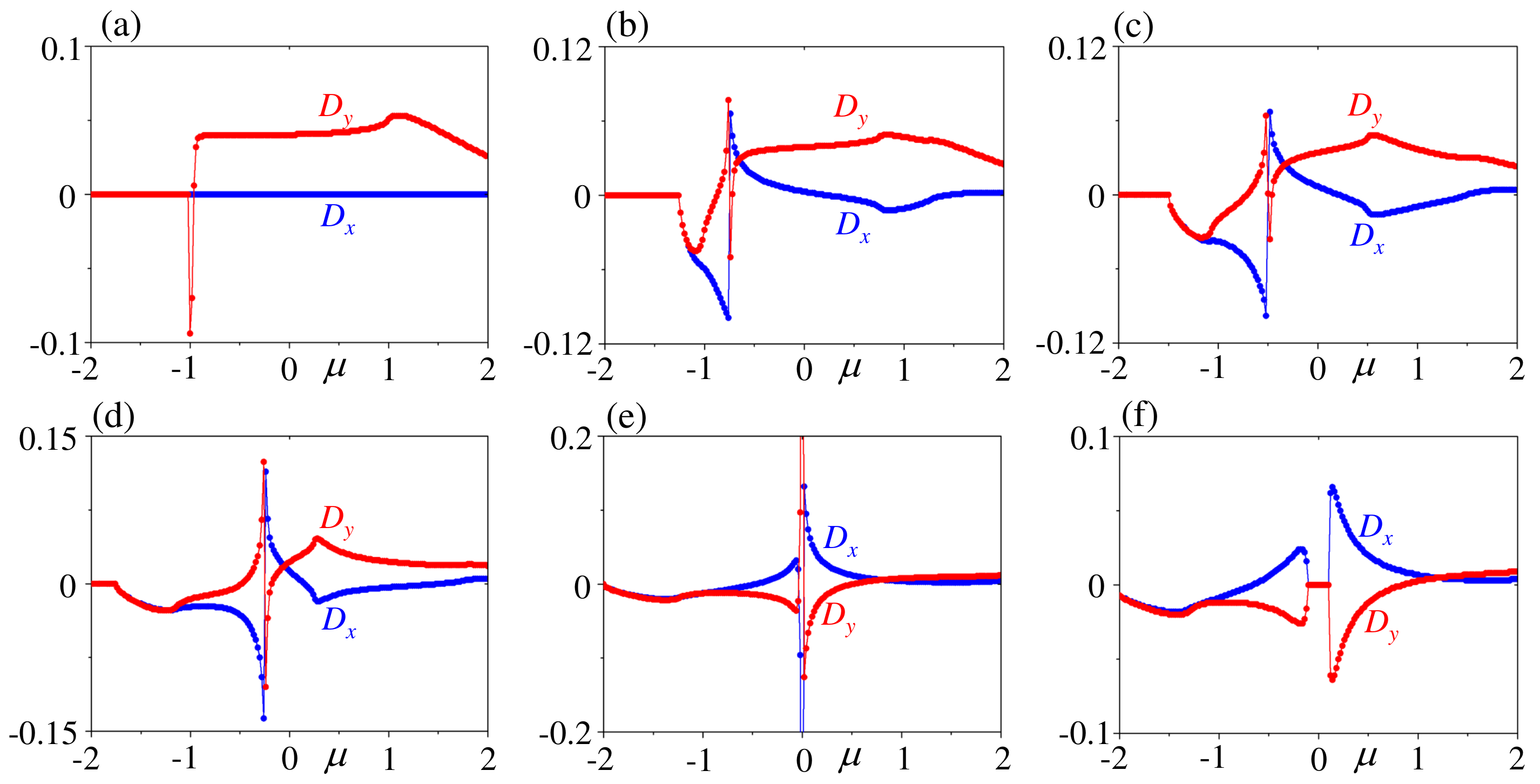}
		\caption{(Color online) The evolutions of $D_{x}$ and $D_{y}$ with respect to the movement and annihilation of DPs. Common parameters
are $t_{0}=t_{1}=0.5$. (a) $\delta=0$, (b) $\delta=0.25$, (c) $\delta=0.5$, (d) $\delta=0.75$, (e) $\delta=1$ and (f) $\delta=1.1$. }
		\label{dxdy}
	\end{center}
\end{figure*}

To see how the topological properties as well as the locations of band degeneracies evolve with
respect to $\delta$, we write down the energy spectra explicitly, which read
\begin{widetext}
\begin{eqnarray}
E_{\pm}(\bk)=2t_{1}\cos k_{x}\cos k_{y}\pm\sqrt{(2t_{1}\sin k_{x}\sin k_{y}+\delta)^{2}+4t_{0}^{2}(\cos k_{x}+\cos k_{y})^{2}}.
\end{eqnarray}
\end{widetext}
It is easy to see that, depending on the sign of $\delta$, the band degeneracies are forced to appear either
on the line $k_{y}=k_{x}+\pi$ or on the line $k_{y}=-k_{x}+\pi$. Without the loss of generality, below
we take $t_{0}=t_{1}=0.5$ and focus on $\delta\geq 0$ for a concrete discussion. For this set of parameters,
one can find that the band degeneracies are located at two inequivalent momenta, $\bQ_{1}=(-\arcsin\sqrt{\delta},-\arcsin\sqrt{\delta}+\pi)$
and $\bQ_{2}=(-\pi+\arcsin\sqrt{\delta},\arcsin\sqrt{\delta})$. Furthermore, the band degeneracies are located
at the energy $E=\delta-1$ when $0\leq\delta\leq1$. Two limits are of special interest. One limit is at $\delta=0$. When $\delta=0$,
$\bQ_{1}$ and $\bQ_{2}$ are equivalent up to a reciprocal lattice vector.  By expanding the Hamiltonian around $(0,\pi)$,
the low-energy Hamiltonian up to the second order in momentum takes the form
\begin{eqnarray}
H_{(0,\pi)}(\bq)=-1+\frac{q^{2}}{2}
-q_{x}q_{y}\sigma_{z}+\frac{1}{2}(q_{x}^{2}-q_{y}^{2})\sigma_{x},
\end{eqnarray}
where $q^{2}=q_{x}^{2}+q_{y}^{2}$. According to this low-energy Hamiltonian, one can find that the dispersions of the two bands follow
$E_{-}(\bq)=-1$ and $E_{+}(\bq)=-1+q^{2}$, which indicates that
the upper dispersive band touches quadratically with the lower flat band
at $(0,\pi)$.  The quadratical touching point has winding number $\pm2$ ($+2$ or $-2$ depends on
the detail of definition), which can be viewed
as the mergence of two DPs with the same winding number $\pm1$. The other limit
is at $\delta=1$. When $\delta=1$,
the two band degeneracies merge again at
$\mathbf{Y}=(-\pi/2,\pi/2)$.  By expanding the Hamiltonian around $\mathbf{Y}$,
however, one will find that the low-energy Hamiltonian up to the second order in momentum
takes the form
\begin{eqnarray}
H_{\mathbf{Y}}(\bq)=-q_{x}q_{y}
+\frac{q^{2}}{2}\sigma_{z}+(q_{x}-q_{y})\sigma_{y}.
\end{eqnarray}
The energy dispersion is quadratic only in the direction $q_{x}=q_{y}$.
This band degeneracy has zero winding number and corresponds to a SDP discussed previously.
It may look counterintuitive at first sight that the winding number is not conserved when $\delta$ is varied
from $0$ to $1$. However, it is worth noting that the winding number is only well-defined for
the low-energy Hamiltonian which has an emerging chiral symmetry when $\delta\leq1$ and neglecting
the $d_{0}$ term which does not affect the geometrical properties of the Bloch wave functions. The
full Hamiltonian itself does not have the chiral symmetry. In addition, even for the low-energy Hamiltonian,
its chiral operator $C$,  which satisfies $\{C,H(\bq)\}=0$, is not fixed.
Instead, it evolves from $C=\sigma_{y}$ to $C=\sigma_{x}$ when $\delta$ is varied from
$0$ to $1$ (a more thorough discussion about this can be found in Ref.\cite{Montambaux2018}). For this two-band Hamiltonian,
the winding number counts the number of times that the vector composed by the remaining  components of $\{d_{x},d_{y},d_{y}\}$
rotates around the axis characterized by the chiral operator
when the momentum changes in a closed path encircling the band degeneracy. As the chiral operator changes,
the non-conservation of winding number is similar to the angular momentum,
which is also not conserved if the rotation symmetry axis is changed.

More details about how the band structure changes with respect to $\delta$ are provided in Fig.\ref{spectrum}.
From Fig.\ref{spectrum}(a), one can see that when $\delta=0$, the lower band is completely flat and the upper dispersive band
touches it quadratically  at the corner of the first Brillouin zone. When $\delta$ becomes nonzero,
the lower band also becomes dispersive, but it remains flat along the line $\mathbf{M}$-$\mathbf{X}$. From Figs.\ref{spectrum}(b)-(d),
one can find that, within the regime $0<\delta<1$,
the two bands touch at two inequivalent momenta and
form two DPs on the line $\mathbf{M}$-$\mathbf{Y}$. It is worth noting that, in this regime,
the tilt of the linear Dirac cones falls into the critical situation which corresponds to
the boundary between type-I and type-II Dirac cones. For linear Dirac cones with such a critical
tilt, they are commonly dubbed as type-III Dirac cones for differentiation. From Fig.\ref{spectrum}(e), one can see
that when $\delta=1$, the
two DPs merge at $\mathbf{Y}$ and form a SDP. Here the SDP is a critical point which separates
the semimetal phase from the insulator phase. When $\delta>1$, the two bands are separated, and
the upper band becomes flat along the line $\mathbf{M}$-$\mathbf{Y}$, as shown in Fig.\ref{spectrum}(f).

Now let us investigate how $D_{x}$ and $D_{y}$ respond to the change of band structure.
Fig.\ref{dxdy}(a) shows the result for $\delta=0$. For this case, we find that $D_{x}$
vanishes identically, which is due to the accidental mirror symmetry
$\mathcal{M}_{y}$ at this limit. In contrast, $D_{y}$ has a dramatic change
when the chemical potential crosses $\mu=-1$. As the quadratical touching point is located at
$E=-1$ when $\delta=0$, the dramatic change in $D_{y}$ at $\mu=-1$ corresponds to that the doping level
meets the quadratical touching point at which the Berry curvature is divergent.
As a finite $\delta$ breaks the mirror symmetry $\mathcal{M}_{y}$, $D_{x}$ also becomes finite when the chemical potential crosses the bands.
Most remarkably, from Figs.\ref{dxdy}(b)-(d), we find that, when $0<\delta<1$, $D_{x}$ and $D_{y}$ simultaneously undergo a dramatic change
when the doping level sweeps $\mu=-1+\delta$. We have discussed previously that the DPs are located at $E=-1+\delta$ when
$0<\delta<1$, so the reason for the dramatic change is also that the doping level meets the DPs at which
the Berry curvature is divergent. When the DPs merge together at $\delta=1$,
$D_{x}$ and $D_{y}$ undergo a dramatic change at $\mu=0$ as expected, but
with their peak values  greatly enhanced, as shown in Fig.\ref{dxdy}(e). When $\delta>1$,
both $D_{x}$ and $D_{y}$ vanish when the doping level is located within the bulk gap,
and their peak values are located near the band edge, as shown in Fig.\ref{dxdy}(f).
Because  the two bands are not symmetric  about $E=0$ due to the existence of the $d_{0}$ term,
$D_{x}$ and $D_{y}$ are also not symmetric about $\mu=0$.

\section{V. Discussion and conclusion}

Although the class AI does not host strong topological insulator phase in two dimensions,
we have shown that the band structures of inversion asymmetric insulators and semimetals belonging
to this class can have nontrivial geometrical and topological properties. For insulators close to
the critical points, we find that the low-energy physics is described by either linear Dirac Hamiltonians
or semi-Dirac Hamiltonians. For both kinds of Hamiltonians, the local Berry curvature near the band edge
will become more prominent with the decrease of Dirac mass, and will change sign when the system
crosses the critical point. Interestingly, we find that the BCD follows the same behavior as the Berry curvature,
indicating that the nonlinear Hall effect can reflect the change in local quantum geometry as well as
the global topology of the band structures in such materials. For the semimetals, in this work we have investigated the Mielke model.
For this model, the distribution of the local Berry curvature and the topological properties of the band degeneracies
will change with the movement of DPs in momentum space. As the local Berry curvature is divergent at the DPs
and the density of states is nonzero for energy at which the type-III DPs are located, we find that
the BCDs also become more and more prominent when the doping level is tuned more and more close to the
DPs. In addition, as the Berry curvatures of conduction and valence bands have opposite signs, we find
that the BCDs  sharply reverse their signs when the doping level is tuned across the DPs.
As two dimensional materials have advantage in tuning the doping level, this remarkable sensitivity
of nonlinear Hall effect to the chemical potential near the DPs may have interesting applications in optoelectronics,
like sensors,
and can also be applied  to determine the locations of DPs in real materials as a complementary method of
angle-resolved photoemission spectroscopy.

From the three models we have studied, we find that the nonlinear Hall effect
can be rather prominent in both doped class AI insulators and semimetals. Our study also
reveals that the tilt of gapped or gapless Dirac cones can benefit the enhancement
of nonlinear Hall effect near the band edge. This enhancement can be understood by noting that the BCD is
co-determined by the density of states, Fermi velocity and Berry curvature
on the Fermi surface. For upright gapped Dirac cones, because the density of states and the Berry curvature
are finite and the Fermi velocity vanishes
at the band edge, the BCD will accordingly vanish when the doping level is exactly located at the band edge. For
upright gapless Dirac cones, while the Berry curvature is divergent and the Fermi velocity is finite,
the density of states vanishes when the Fermi level is exactly located  at the DPs, also leading a
zero BCD. Without affecting the Berry curvature, a
finite tilt of gapped Dirac cones can make the Fermi velocity at the band edge finite,
and a sufficient  strong tilt of gapless Dirac cones can make the density of states nonzero even
though the Fermi level is exactly located at the DPs, consequently leading to the enhancement.

Similar to the linear anomalous Hall effect, nonlinear Hall effect in fact can also have contributions from
disorder-induced side jump and skew scattering~\cite{Du2019b,Konig2019HE,Xiao2019NHE,Nandy2019NHE,Isobe2020,Du2020NHE}.
However, In this work we have restricted ourselves to the intrinsic part.
We will leave the investigation of the disorder-induced contributions in this class of materials
for future work.
Overall, from a low-energy perspective, the three models we have studied are quite representative in the description
of class AI materials without inversion symmetry. As the class AI requires the
spin-orbit coupling to be negligible, our findings in this work are relevant to materials
consisting of only light elements. Among various possibilities, we suggest the application
of organic materials to test our predictions since very recently the experimental observation of nonlinear Hall
effect in such materials has been reported~\cite{Kiswandhi2021NHE}.

\section{VI. Acknowlegements}

Z.S.L. and H.H.Z. are supported by the National Natural Science Foundation of China (NSFC) under Grant No. 11875327.
Z.Y. is supported by the National Science Foundation of
China (Grant No. 11904417) and Z.Y. is supported by the National Natural Science Foundation of China (Grant No.11904417) and the Natural Science Foundation of Guangdong Province (Grant No. 2021B1515020026).

\bibliography{dirac}

\end{document}